\documentclass[12pt,preprint]{aastex}
\begin{document}
\title{Evolution of Mid--IR Excess Around Sun--like Stars: \\
Constraints on Models of Terrestrial Planet Formation} 
\author{M.R. Meyer\altaffilmark{1},
J.M. Carpenter\altaffilmark{2},
E.E. Mamajek\altaffilmark{3},
L.A. Hillenbrand\altaffilmark{2},
D. Hollenbach\altaffilmark{4},
A. Moro--Martin\altaffilmark{5},
J.S. Kim\altaffilmark{1},
M.D. Silverstone\altaffilmark{1},
J. Najita\altaffilmark{6},
D.C. Hines\altaffilmark{7},
I. Pascucci\altaffilmark{1}, 
J.R. Stauffer\altaffilmark{8},
J. Bouwman\altaffilmark{9}, \&
D.E. Backman\altaffilmark{10}} 
\altaffiltext{1}{Steward Observatory, The University of Arizona, Tucson, AZ; mmeyer, serena, msilverstone, pascucci -- @as.arizona.edu}
\altaffiltext{2}{Astronomy, California Institute of Technology, Pasadena, CA; jmc, lah -- @astro.caltech.edu}
\altaffiltext{3}{Harvard--Smithsonian CfA, Cambridge, MA; emamajek@cfa.harvard.edu} 
\altaffiltext{4}{NASA--Ames Research Center, Moffett Field, CA; hollenbach@ism.arc.nasa.gov}
\altaffiltext{5}{Astronomy, Princeton University, Princeton, NJ; amaya@astro.princeton.edu} 
\altaffiltext{6}{National Optical Astronomy Observatory, Tucson, AZ; jnajita@noao.edu}
\altaffiltext{7}{Space Science Institute, Boulder, CO; hines@spacescience.org} 
\altaffiltext{8}{Spitzer Science Center, Pasadena, CA; stauffer@ipac.caltech.edu}
\altaffiltext{9}{Max--Planck Institut f\"ur Astronomie, Heidelberg, Germany; bouwman@mpia.mpg.de}
\altaffiltext{10}{SETI Institute, Mountain View, CA; dbackman@mail.arc.nasa.gov} 

\begin{abstract} We report observations from the Spitzer Space
Telescope (SST) regarding the frequency of 24 $\mu$m excess emission toward
sun-like stars.
Our unbiased sample is comprised of 309 stars with masses 0.7-2.2
M$_{\odot}$ and ages from $<$3 Myr to $>$3 Gyr that lack excess
emission at wavelengths $\leq$8 $\mu$m.  We identify 30 stars that
exhibit clear evidence of excess emission from the observed
$24/8 \mu$m flux ratio.  The implied 24 $\mu$m excesses of these
candidate debris disk systems range from 13 \% (the minimum detectable) to more
than 100\% compared to the expected photospheric emission.  
The frequency of systems with evidence for dust debris emitting
at 24 $\mu$m ranges from 8.5--19 \% at ages $<$300 Myr to $<$ 4 \% 
for older stars.  The results suggest that many, perhaps most,
sun-like stars might form terrestrial planets.
\end{abstract}


\section {Introduction}

Are planetary systems like our own common or rare in the Milky Way
galaxy? The answer depends on what aspect of our planetary system one
is investigating. Gas and dust rich circumstellar disks appear to be a
common outcome of the star formation process \citep{Strom93}. Gas
giant planets within 5 AU (presumably formed from these disks)
surround $>$6\% of sun-like stars \citep{Marcy05} while the detection
of terrestrial planets is still in its infancy \citep{Beaulieu06}.
Although debates concerning theories of giant planet
formation continue \citep{Durisen07,Lissauer07}, there
is some consensus regarding the formation of terrestrial planets
\citep[e.g.][]{Nagasawa07}. Starting with a swarm of 1
km-sized planetesimals, orderly growth of larger bodies proceeds
rapidly ($<$ 1 Myr) out to at least 2 AU. When the gravitational cross
section greatly exceeds the geometrical cross section of the largest objects,
growth transitions from orderly to oligarchic with the biggest bodies
growing fastest in a runaway process. The final stage, chaotic growth,
is characterized by high velocity collisions between the few remaining
large bodies in the system.  Remaining challenges include the
formation of km-sized planetesimals in the face of gas drag on
meter-sized bodies \citep{Weidenschilling77} and Type I migration of
lunar-mass objects in a remnant gas disk \citep{Nelson05}.

Yet there are few observational tests of this developed theory. The
physical characteristics of the terrestrial planets, their satellites,
and the asteroid belt provide constraints on the formation of our
solar system \citep{Bottke05,O'Brien06}.  Observations of
circumstellar dust debris surrounding sun-like stars can be used to
trace the presence of planetesimal belts of larger parent bodies
\citep{Meyer07} and thus constrain theories of planet formation. 
Far--IR observations at 70 $\mu$m with the
Spitzer Space Telescope suggest that 10-15\% of sun-like stars
possess cool outer dust disks that are massive analogs of the Kuiper
Belt \citep{Bryden06}.  Yet few mature stars exhibit mid--infrared
excess indicative of terrestrial temperature material
\citep{Beichman05}.  Recent work with the Spitzer Space
Telescope has begun to assess the frequency of this emission around
younger stars \citep{Chen05, Hernandez06}.  
In this contribution we use
Spitzer data to investigate the frequency of mid-IR excess emission,
which may originate from 1--10 AU, observed toward sun--like stars 
over a wide range of ages spanning the epoch of terrestrial
planet formation in our Solar System.

\section{Observations} 

Observations were obtained as part of the Formation and Evolution of
Planetary Systems (FEPS) Legacy Science Program \citep{Meyer06}. Our
sample consists of 328 ``sun-like'' stars with spectral types F5-K3
and masses ranging from 0.7-2.2 M$_{\odot}$ (though strongly peaked at
1.0 M$_{\odot}$). The sample was constructed so that roughly
equal numbers of stars were selected in logarithmically spaced
age-bins from 3 Myr to 3 Gyr (each bin spanning a factor of x3 in
age). Stars $<$ 100 Myr were largely drawn from young stellar
populations within the Local Association, often 
members of OB and T Associations. Ages for
these stars were estimated from pre-main sequence evolutionary tracks,
as well as kinematic association with groups of known age
\citep[e.g.][]{Mamajek02}. Older main sequence stars were selected
from a volume-limited sample of stars taken from the HIPPARCOS
catalog. Ages for these stars were estimated from calcium H \& K
emission-line indices which trace stellar activity levels using
the calibration of \citet{Donahue98}. Errors in age for both the
young stellar populations as well as the main sequence stars are
estimated to be $<$ 50 \%, though uncertainties in the absolute
calibration of these ages are not well
understood.  A fraction of our sample in the age
range from 30 Myr to 1 Gyr were selected to be members of open
clusters with well determined ages \citep[e.g.][]{Stauffer05}. 
Our sample selection is described in \citet{Meyer06} and details concerning the
age estimates for each star are given in Hillenbrand et al.  (in
prep). We fit Kurucz model atmospheres to B/V (Tycho) photometry from
HIPPARCOS and JHK$_S$ photometry from 2MASS. We assumed solar
metalicity and surface gravities estimated from the position of each
star in the H--R diagram, performing a non-linear least squares fit
for T$_{eff}$ and solid angle. For stars within 75 pc, we
assumed A$_v$=0 while for more distant targets A$_v$ 
was a free parameter. 

All sources were detected at 8 $\mu$m and 24 $\mu$m with SNR $>$
30 using the SST.  Photometry at 8 $\mu$m was derived from sub--array
observations with the IRAC instrument \citep{Fazio04}.  We began our
analysis with data processed through the S13 pipeline.  Cosmic ray
rejection was implemented and corrections for spatially-dependent
pixel area and filter response variations were applied.  Aperture
photometry was derived from from a 3.7'' radius aperture 
(with sky anuli ranging from 12.2--24.4'') using 
a modified version of IDLPHOT and placed
on the standard flux scale recommended by \citet{Reach05}.  Systematic
calibration errors are estimated to be $<$ 2 \% in each band.
Random photometric uncertainties were estimated from the repeatability
of 64 observations obtained at each dither position for each source
resulting in minimum uncertainties of 1\% at $8 \mu$m.
Photometry at 24 $\mu$m was derived from either 28 or 56 exposures
with integration times of 3 or 10 seconds each with the MIPS
instrument \citep{Rieke04}. We began with S13 pipeline data and
photometry was derived using the MOPEX software. 
Fluxes were estimated from a PSF--fitting algorithm and placed
on the flux scale recommended by the SSC.  Calibration errors are
thought to be $<$ 4\% \citep[cf.][]{Engelbracht07}.  The minimum
random uncertainties in the 24 $\mu$m photometry are 1 \%.

\section{Analysis}

Of the parent sample of 328, 14 stars were selected for our initial
IRS search for remnant circumstellar gas \citep{Pascucci06} on the
basis of previously detected dust disk signatures.  Our analysis
starts with the unbiased FEPS sample of 314 (= 328 - 14) stars
spanning a range of ages from $<$3 Myr to $>$3 Gyr.  We use the
$24/8 \mu$m flux ratio to search for stars that exhibit excess
emission.  These data are plotted in Figure 1 as a function of $8 \mu$m 
flux.  Note that the brighter sources tend to be nearby
(older) field stars, while fainter targets tend to be more distant
(and younger) sources.  Several systems exhibit flux
ratios indicative of a 24 $\mu$m excess. 
The expected photospheric ratio ($24/8 \mu$m) in this
diagram is approximately 0.116.  In
order to define an empirical ``blue envelope'' of stars without excess, 
we employed a sigma--clipping algorithm to the distribution of
$24/8 \mu$m flux ratios.  Initially, the mean flux ratio was computed
to be 0.190 with $\sigma$ = 0.82, and two outliers with ratios
beyond 3$\sigma$ (indicating obvious excess).  
Removing these two outliers, we recompute the mean and
sigma, resulting in identification of three additional sources with
smaller excesses.  Repeating this process a total of seven times, the
values converge with a mean of
0.117 and $\sigma$ = 0.005 as shown in Figure 1, 
consistent with model predictions. 
We identify 35 (positive) outliers which we attribute to excess
emission in the $24 \mu$m band.  Of these, only five exhibit excess emission 
at $\leq 8 \mu$m and were previously identified in \citet{Silverstone06} as 
optically--thick primordial gas rich disks.  Because we are 
interested in understanding the transition to debris disks at 24
$\mu$m we remove these five from the sample of excess stars under
consideration ($314-5=309$).  Assuming that the color excesses observed 
are due to excess emission at 24 $\mu$m, our 3$\sigma$ detection limit of
($0.117 + 3 \times 0.005 = 0.132$) corresponds to a 13 \%
excess at 24 $\mu$m 
compared to the expected photospheric emission ($0.132/0.117-1$).
The largest inferred 24 $\mu$m excess 
is just over 100 \%
compared to the photosphere ($0.245/0.117-1$).  The 30 sources with
detectable excess from our sample of 309 are listed in Table 1 as a
function of age.
According to the Shapiro--Wilk test, the distribution of $24/8 \mu$m flux 
ratios for the 279 sources {\it without} excess is not gaussian 
(P $<$ 0.01 \%).  
We tested to see whether the mean ratio
of 24/8 $\mu$m emission was a function of source brightness.  
For the 140 targets with $8 \mu$m $>$ 128 mJy the mean was 0.1137 with 
$\sigma$ = 0.0036, while for the 139 fainter 
targets the mean was 0.1199 with $\sigma$ = 0.0047. 
This small offset could be due to uncertainties in flux 
calibration as a function of integration time (Carpenter et al. 2007; 
Engelbracht et al. 2007)\footnote{Adopting these offsets in the mean
flux ratio (and associated 3 $\sigma$ limits)
would result in identification of one new excess object
(HD 43989, 30--100 Myr old) and removal of [PZ99] 161618.0-233947
from Table 1.}.
As a result, the errors quoted on the reported 
excesses include the random errors in the 
$24/8 \mu$m ratio (typically 1--2 \%), 
as well as the dispersion in our estimate of the 
photospheric color (4.3 \%) rather than the error in the mean, 
added in quadrature. 

In Figure 2, we present the fraction of stars exhibiting 24 $\mu$m
excess emission in our sample as a function of age.  Each bin spans a
factor of 3 in age.  The errors in the ordinate are Poisson, computed
following \citet{Gehrels86} with excess fractions as follows: (5/30) for
stars 3--10 Myr, (9/48) for 10--30 Myr, (5/59) for 30--100 Myr, 
(9/62) for 100--300 Myr, (2/53) 300--1000 Myr, and (0/57) for 
stars 1--3 Gyr old.  The KS test suggests that the distributions of
$24/8 \mu$m flux ratios (Fig. 1) for the sample $<$ 300 Myr (N = 199)
and those $>$ 300 Myr (N = 110) are inconsistent with having been
drawn from the same parent population (P $<$ 10$^{-10}$).  We note
that the errors in age quoted above act to diffuse sources to younger
as well as older ages.  Because there are more excess objects in younger
bins, errors in age tend to increase the excess fractions at older
ages.  As a result, the abrupt drop in the excess fraction at 300 Myr 
may be even more dramatic than detected here. 

\section{Discussion}

We associate the observed 24 $\mu$m excess with dust debris generated through
collisions of planetesimals.  One of our excess stars identified in
Table I (HD 12039) was studied in detail by \citet{Hines06}.  Models
of this debris disk (with fractional 24 
$\mu$m excess $0.151/0.117 - 1 = 0.29$) 
suggested a dust mass of $\sim$ 10$^{-5}$ M$_{\earth}$
located between 4--6 AU.  
The magnitude of all our detected 24 $\mu$m excesses are 
within a a factor of $\times$ 3 (relative to the photosphere) 
compared to HD 12039. 
Results similar to ours, 
have been reported for samples of FGK stars in open clusters
\citep{Gorlova06,Siegler07}.  Our sample is comprised
of 60 open cluster stars with discrete ages of 55 (5 members of IC
2602), 90 (13 members of $\alpha$ Per), 110 (20 Pleiades), and 600 Myr
(22 Hyades), as well 249 field stars.
We have analyzed the statistics for sub-samples where they
overlap. While the excess fractions for open clusters with ages
30--100 Myr and 100--300 Myr are {\it greater} than the
comparable field star samples($3/18$ vs. $2/41$ and $5/20$ vs. $4/42$ 
respectively), the results are formally consistent
with each other. This suggests that there is no strong dependence of
debris disk evolution on star--forming environment, though larger 
samples could reveal a difference.

A key question is whether stars observed to have excess
at 24 $\mu$m in one age bin are the same cohort of stars with 24 $\mu$m excess
in another. In other words, do the same 10--20 \% of
sun--like stars with excess evolve from one age bin to the next with a
constant fraction; or are they distinct groups of stars, that persist
in the observed state for a short time?
Our observations trace excess emission from 21.7--26.4 
$\mu$m toward stars lacking excess emission $\leq 
10 \mu$m (corresponding to a lack of dust generating 
planetesimals inside 1 AU \citep{Silverstone06}).  
Assuming blackbody emission from large grains implies 
dust at radii from $\sim$ 4--7 AU.  
Maximum dust production during the evolution of a 
planetesimal swarm is thought to 
occur between runaway and chaotic growth when the largest planetesimals 
reach $\sim$ 2000 km at a given radius \citep{Kenyon04,Kenyon06}.  
The timescale for this 
goes as $\tau \sim a^{1.5} \sigma_{disk}^{-1}$ 
where a is the orbital radius and $\sigma_{disk}$ 
is the mass surface density of solids
in the disk (\citet{Goldreich04}).  
Assuming $\sigma_{disk} \sim \sigma_{o} a^{-p}$, and that
$0 < p < 1$ (\citet{Kitamura02}), 
a range of $\times 2$ in radius
corresponds to $\times 3-6$ in time.  Thus 
the emission we observe might not persist over timescales 
much larger than our age bins
\footnote{While the published Kenyon and Bromley models 
predict the duration of $24 \mu$m excess emission $> \times 3-6$ 
in time, they also predict hot dust at smaller radii
covering a wider range of radii than our observations imply.}.
Perhaps many stars go through this phase of 24 $\mu$m excess, but at
different times.  
A range of $\times$ 100 in initial mass surface density 
(\citet{Andrews05}) could translate
into a range of $\times$ 100 in evolutionary timescales.   If so, 
one might expect smaller excesses at later times (produced by 
lower mass disks).  In comparing the mean detected excess for stars 
3--30 Myr old (0.359 with $\sigma$ = 0.199) with that for 
stars 30--300 Myr old (0.345 with $\sigma$ = 0.116), 
we find no evidence that this is the case
(though the samples are dominated by stars lacking detectable excess). 
Nevertheless, one might consider {\it summing} the fractions of stars with 
24 $\mu$m excess between 3-300 Myr, resulting in an overall fraction of stars
with evidence for terrestrial planet formation greater than 60 \%!
Averaging the results over factors of ten in age results in excess
fractions of 18, 12, and 2 \% at ages 3--30 Myr, 30--300 Myr, and
0.3--3 Gyr, implying at least 32 \% of sun--like stars 
exhibit evidence for terrestrial 
planet formation (provided that the epoch of 24 $\mu$m excess
emission lasts $\leq \times 10$ in age).
We note that in this scenario, the planets formed later from lower
mass disks will be smaller (\citet{Kenyon06}).

Results to date suggest that: a) primordial disks between 0.3-3 AU
dissipate or agglomerate into larger bodies on timescales comparable
to the cessation of accretion\citep{Haisch01}; and b) few stars harbor optically-thin
inner disks between 3-30 Myr \citep{Silverstone06}.  Based on theoretical
considerations, we expect that planetesimals belts evolved rapidly 
within 3 AU. 
\citet{Rieke05} explore the evolution of 24 $\mu$m excess emission
around a sample of A stars observed with SST and IRAS.  They
deduce a characteristic timescale of 150 Myr for strong excesses 
to decay.  However, care must be taken in comparing these results
to ours as: 1) the dust masses detected by Rieke 
et al. are likely {\it larger} than the dust masses detected here;
and 2) similar temperature dust traces
distinct radii for stars of different luminosity. In general, 
the fractional 24 $\mu$m excesses around A stars are larger 
than around G stars.
The observed
duration of the excess phase for both samples is 
longer than expected if the emission results solely from dust
production well inside 10 AU.  An important caveat to our results
is that we have not assessed whether the 24
$\mu$m excesses we have detected are tracing warm dust in the
terrestrial planet zone, or the Wien--side of the Planck function from
cooler dust.  If, at 24 $\mu$m, we are
seeing cooler dust generated at radii
beyond 10 AU, we would expect to observe it at later times.  We
also note that the maximum excess ratios predicted by Kenyon and
Bromley are larger than the excesses detected here.

\citet{Wyatt07} provide steady--state models of warm dust production
around sun--like stars.  On the basis of comparing the observed IR
luminosity of several systems to these models as a function of age,
they conclude that most (5/8) of the stars exhibiting evidence for
warm dust must be in a transient state of evolution and not
participating in a steady--state collisional cascade.  However, the
three systems  
with ages $<$ 300 Myr (one of which is HD 12039) 
can be explained with equilibrium
models.  
Global analysis of the SED for all sources identified here
will be required to directly compare the evolutionary state of these
systems with the models (Carpenter et al., in preparation). 

More work is needed to define the transition from primordial to debris
disk \citep[e.g.][]{Padgett06}.  Given the short time expected for
collisional evolution of inner planetesimals belts ($<$ 3 Myr), and
the 1-10 Myr lifetime of primordial disks, it may be
difficult to detect the onset of collisional evolution.  We
suggest that SST observations at 24 $\mu$m can be interpreted as evidence
for terrestrial planet formation occuring around many (19--32 \%), 
if not most (62 \%), sun--like stars.  This range is higher than 
the observed frequency of gas giant planets 
(6.6--12 \% within 5--20 AU \citet{Marcy05})
but comparable to the inference that cool dust debris beyond 
10 AU might be very common \citet{Bryden06}.  Radial
velocity monitoring of low mass stars, micro-lensing surveys, as well
as transit surveys such as COROT and Kepler, will provide
critical tests of our interpretation. 

We thank all members of the FEPS, IRAC, MIPS, and SSC teams
for their efforts, as well as Scott Kenyon, Nick Siegler,
and George Rieke for valuable discussions, and an anonymous referee
for helpful suggestions.  This work was based
on observations made with the Spitzer Space Telescope, which is 
operated by the Jet Propulsion Laboratory, California Institute 
of Technology under a contract with NASA.  Support for this work
was provided by NASA through an award issued by JPL/Caltech. 


\begin{deluxetable}{lcccrcc}
\tabletypesize{\scriptsize}
\tablewidth{0pt}
\tablecolumns{9}
\setlength{\tabcolsep}{0.06in}
\tablecaption{Systems with MIPS--24 $\mu$m Excess \label{properties}}
\tablehead{
\colhead{Source} & \colhead{Dist} & \colhead{log(age)} &
\colhead{T$_{eff}$} & \colhead{log(L$_{\star}$/L$_\odot$)} &
\colhead{f$_{24 \mu m}$(excess)/f$_{24\mu m}$}(phot) & \colhead{$\sigma$} \\
 & (pc) & (yr) & (K) & (dex) & & }
\startdata
  1RXS J051111.1+281353     & 199 & 6.5-7   & 5270 &  0.71 & 0.239 & 0.045\\
        RX J1600.6-2159     & 161 & 6.5-7   & 5330 &  0.27 & 0.190 & 0.045\\
$[$PZ99$]$ J161459.2-275023 & 114 & 6.5-7   & 5500 & -0.14 & 0.628 & 0.047\\
$[$PZ99$]$ J155847.8-175800 & 161 & 6.5-7   & 4660 &  0.20 & 0.479 & 0.046\\
$[$PZ99$]$ J161618.0-233947 & 161 & 6.5-7   & 5250 &  0.22 & 0.141 & 0.045\\
               HD 22179     &  68 & 7-7.5   & 5990 &  0.02 & 0.336 & 0.045\\
              HD 116099     & 140 & 7-7.5   & 6010 &  0.19 & 0.188 & 0.045\\
              HD 141943     &  67 & 7-7.5   & 5810 &  0.43 & 0.210 & 0.045\\
              HD 281691     &  73 & 7-7.5   & 5140 & -0.42 & 0.156 & 0.045\\
                  MML 8     & 108 & 7-7.5   & 5810 &  0.15 & 0.737 & 0.047\\
                 MML 17     & 124 & 7-7.5   & 6000 &  0.43 & 0.612 & 0.046\\
                 MML 28     & 108 & 7-7.5   & 5000 & -0.35 & 0.413 & 0.046\\
                 MML 36     &  98 & 7-7.5   & 5270 &  0.03 & 0.541 & 0.046\\
                 MML 43     & 132 & 7-7.5   & 5410 &  0.06 & 0.154 & 0.045\\
                 HD 377     &  39 & 7.5-8   & 5850 &  0.09 & 0.332 & 0.045\\
               HD 12039     &  42 & 7.5-8   & 5690 & -0.05 & 0.287 & 0.045\\
                 HE 750     & 176 & 7.5-8   & 6360 &  0.28 & 0.197 & 0.046\\
                 HE 848     & 176 & 7.5-8   & 6310 &  0.47 & 0.537 & 0.047\\
                    W79     & 152 & 7.5-8   & 5380 & -0.29 & 0.242 & 0.046\\
               HD 19668     &  40 & 8-8.5   & 5420 & -0.23 & 0.192 & 0.045\\
               HD 61005     &  35 & 8-8.5   & 5460 & -0.26 & 1.096 & 0.049\\
               HD 72687     &  46 & 8-8.5   &  --  & -0.05 & 0.208 & 0.046\\
              HD 107146     &  28 & 8-8.5   & 5860 &  0.02 & 0.329 & 0.045\\
                HII 152     & 133 & 8-8.5   & 5700 & -0.10 & 0.366 & 0.046\\
                HII 250     & 133 & 8-8.5   & 5770 & -0.04 & 0.146 & 0.046\\
                HII 514     & 133 & 8-8.5   & 5720 &  0.04 & 0.250 & 0.046\\
               HII 1101     & 133 & 8-8.5   & 6070 &  0.08 & 0.545 & 0.046\\
               HII 1200     & 133 & 8-8.5   & 6210 &  0.35 & 0.170 & 0.045\\
               HD 85301     &  32 & 8.5-9   & 5600 & -0.14 & 0.343 & 0.046\\
              HD 219498     &  62 & 8.5-9   & 5670 & -0.07 & 0.277 & 0.045\\
\enddata
\end{deluxetable}

\clearpage

\begin{figure}
\plotone{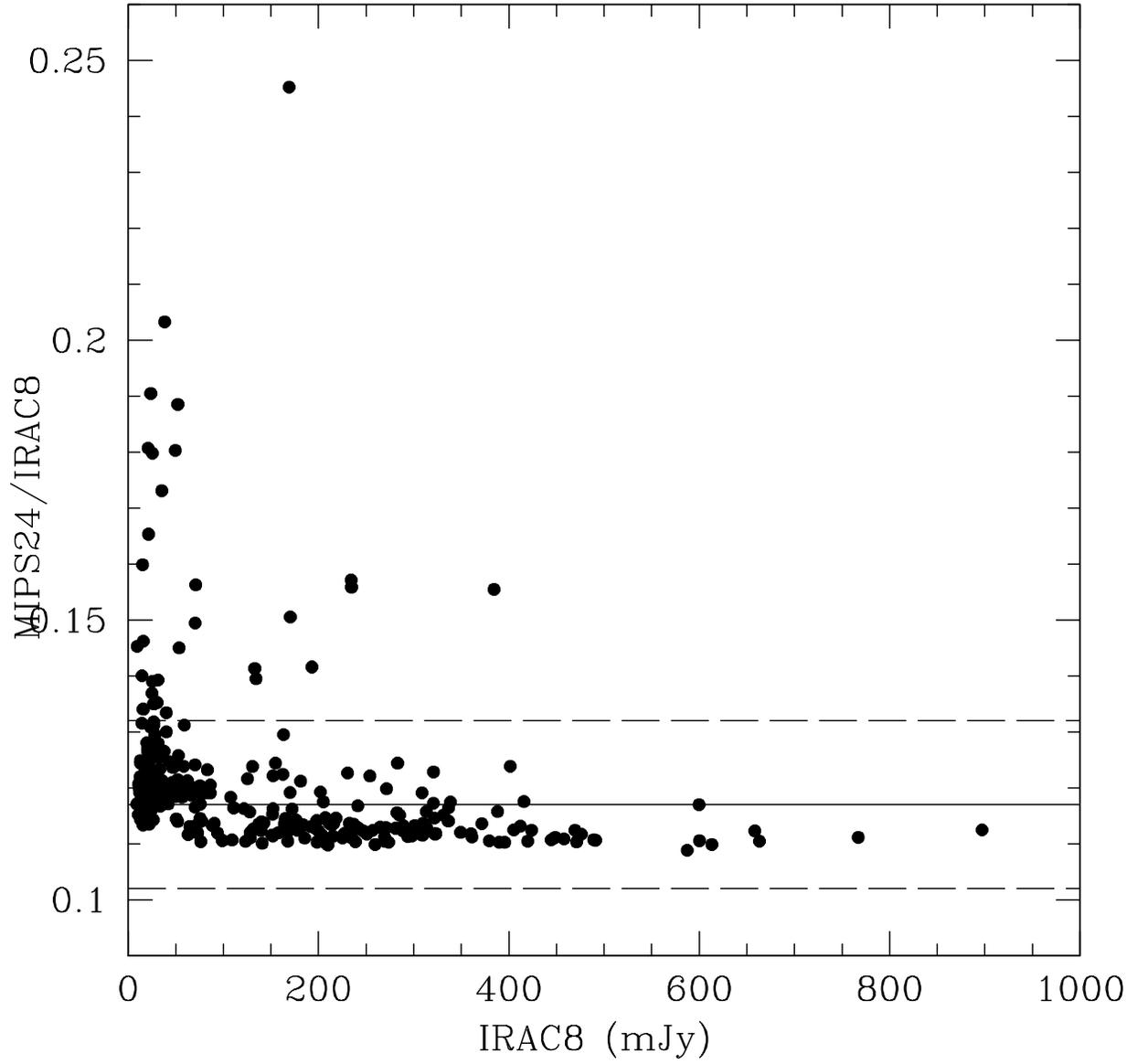}
\caption{$24/8 \mu$m flux ratio plotted as a function of $8 \mu$m 
flux for 314 stars drawn from the unbiased FEPS sample.
The sample mean (solid line) and 3$\sigma$ limits (dashed lines) as described
in the text are shown.}
\label{CMD}
\end{figure}

\vskip 2.0in

\begin{figure}
\plotone{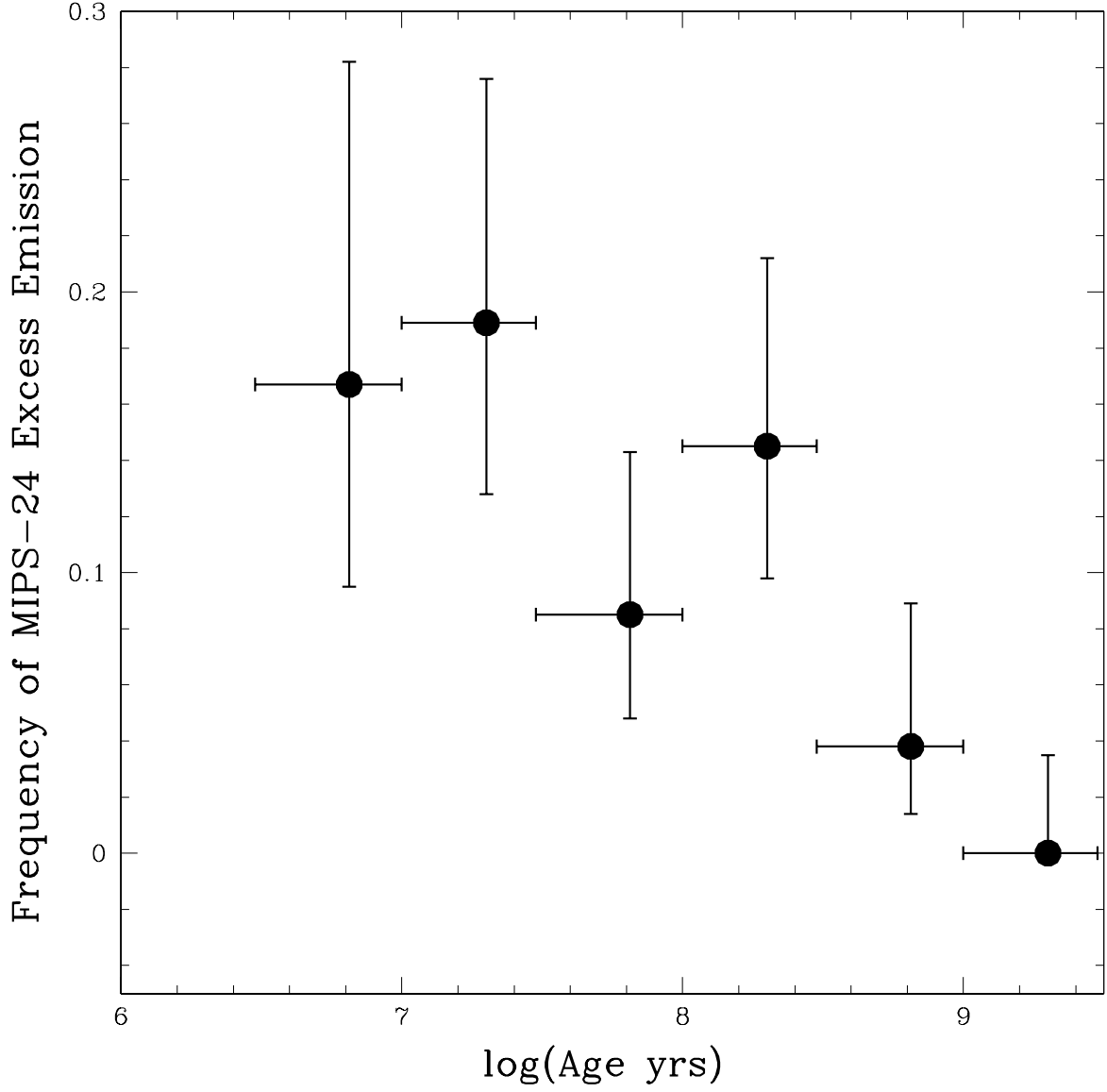}
\caption{The fraction of stars in the sample with detectable $24 \mu$m excess
plotted as a function of age.}
\label{evolution}
\end{figure}

\end{document}